\documentclass[9pt,twocolumn,twoside]{opticajnl}
\journal{opticajournal} 

\usepackage{listings}
\usepackage{xcolor}
\setboolean{shortarticle}{false}

\usepackage{lineno}

\title{Bidirectional schlieren imaging using dual-LED illumination and a lens aperture as a knife-edge}
\author[1]{Vimod Kumar}
\author[1,*]{Manish Kumar}

\affil[1]{Centre for Sensors, Instrumentation and Cyber-Physical System Engineering (SeNSE), Indian Institute of Technology Delhi,\newline Hauz Khas, New Delhi 110016, India}

\affil[*]{kmanish@iitd.ac.in}

\begin{abstract}
In this letter, we demonstrate that employing the Lens Aperture as a Knife-Edge (LAKE) with two identical LEDs has the distinct advantage of obtaining schlieren images along two cutoff orientations. It is possible to control LED brightness to obtain a desirable weighted sum of both horizontal and vertical schlieren cutoff images on each camera frames. Powering the LED sources by an affordable and programmable Arduino board enables a rapid switch between the horizontal and vertical schlieren cutoff that is exactly synced to the camera frames, resulting in crosstalk-free schlieren image sequence.  Furthermore, with a little postprocessing, this monochrome image series can be easily transformed into a colorful sequence of bidirectional schlieren images. 
\end{abstract}

\setboolean{displaycopyright}{false} 

\begin{document}

\maketitle

\section{Introduction}
\begin{figure}[!b]
\centering
\includegraphics[width=0.9\linewidth]{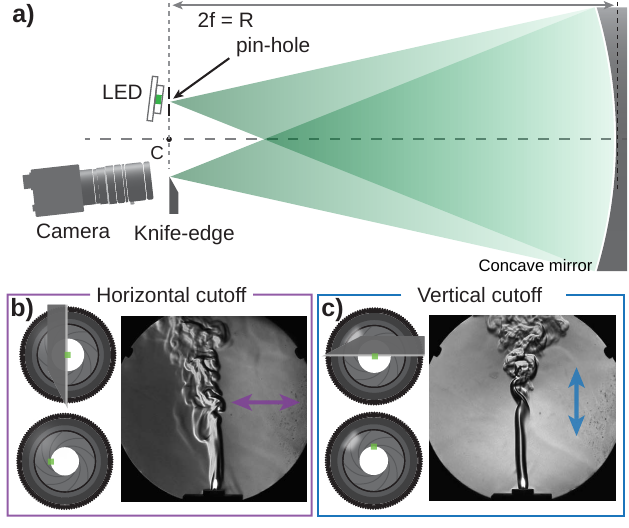}
\caption{Directional nature of schlieren sensitivity. a) Schematic of a standard single-mirror schlieren imaging setup. b) A horizontal schlieren sensitivity via a vertical knife edge cutoff or the vertical edge of the circular lens aperture as a cutoff element. c) A horizontal knife edge or the horizontal edge of the circular lens aperture serves as a cutoff element to achieve vertical schlieren sensitivity. Flow sample: candle plume, Arrows: schlieren sensitivity orientations.}
\label{fig:1}
\end{figure}
Schlieren imaging is a well-known method to visualize flow in transparent media. Traditional schlieren employs a knife edge to translate a density gradient in the flow into an intensity variation on the image plane \cite{topler_optische_1867}. However, this density or refractive index gradient visualization is directionally selective. Only the refractive index gradients that deflect light perpendicular to the knife-edge orientation are visible. The refractive index gradient parallel to the knife edge is ignored \cite{topler_optische_1867, settles2001schlieren}. Figure \ref{fig:1} shows the effect of cutoff orientation on the corresponding density gradient visualization. For a complete refractive index gradient visualization, both horizontal and vertical cutoff are essential. Towards this goal, multiple schlieren techniques have been presented, each with a bi- or multi-directional cutoff. These approaches include the simultaneous use of horizontal and vertical knife-edges \cite{barry1948improved}, square apertures for cutoff with two coplanar bilateral slits sources \cite{edmonson1952radial}, L-shaped knife-edges \cite{speak1954optical}, and color filter enabled color schlieren approaches \cite{Settles1970DirectionIndicating}. In recent years, several advanced approaches to bidirectional schlieren imaging have been developed. Some of these include a calibrated color schlieren using a square source and color filters \cite{Watt2004Theory}, a modified Mach-Zehnder interferometry setup combined with color filters \cite{kleine2006simultaneous}, bidirectional color filters \cite{stricker2006bidirectional}, a Mach-Zehnder-type double path double knife-edge setup \cite{chaloupka2014color}, and very recently an approach having an RGB LED combined with color filters and a color camera \cite{martinez2022simultaneous, martinez2022horizontally}. The two-path bidirectional schlieren approach \cite{barry1948improved,chaloupka2014color} is unique in that it uses no color filters and can produce colored bidirectional schlieren images with a monochrome camera sensor. Regardless of the distinctions between multiple bidirectional schlieren techniques, something is common to all. All existing bidirectional schlieren techniques use a single light source that is either reshaped with slits or masks or split into two pathways with a beam splitter. It is also obvious that all bidirectional schlieren techniques utilize two (or more) knife edges.

In this letter, we describe a bidirectional schlieren approach that does not rely on a single light source or several discrete knife-edges for cutoffs. We employ a dual-LED source-based illumination setup, with one LED responsible for horizontal and the other for vertical cutoff. To the best of our knowledge, no previous research has used independently drivable split light sources for bidirectional schlieren imaging. Our solution also takes advantage of the LAKE approach that deploys a circular internal aperture of the camera lens as a cutoff element \cite{kumar2026using, saxena2026water}. Using this method, we can control the relative strengths of horizontal and vertical cutoffs while also utilizing perfect digital synchronization between camera frames and LEDs to perform crosstalk-free alternate horizontal and vertical cutoffs that can be combined to create colored bidirectional schlieren images. Our dual-LED illumination is driven with a basic programmable Arduino UNO development board. We achieve an imaging speed of 500 fps, currently constrained by the available camera hardware. The LED brightness is sufficient to support a much faster acquisition speed using a high-speed camera. 

\begin{figure}[!t]
\centering
\includegraphics[width=0.95\linewidth]{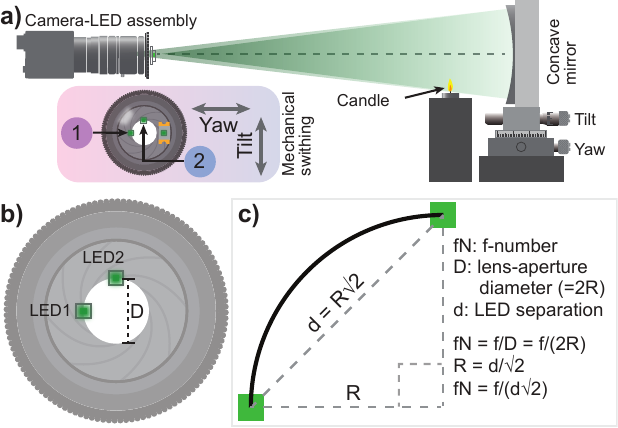}
\caption{Switching between horizontal and vertical cutoff with LAKE schlieren approach. a) A simple tilt and yaw adjustment in the LAKE method allows for switching between horizontal and vertical schlieren sensitivities. The inset depicts an en face view of the camera lens with the physical LED and its corresponding image positions for switching sensitivity orientation. b)-c) Schematics showing the relationship between lens aperture size and LED image position separation for horizontal and vertical schlieren sensitivity.}
\label{fig:2}
\end{figure}

\section{Theory}
The LAKE method has recently evolved as an effective alternative to using an external knife edge as a schlieren cutoff element. Figure \ref{fig:1} illustrates how to obtain a horizontal and vertical cutoff using the lens aperture as a knife-edge. This is consistent since, with point-like light sources, the lens apertures are substantially larger, with the edge of the aperture approximating a straight knife-edge. Unlike older lens designs, new fixed-focal-length camera lenses have fairly circular apertures in a wide range of F-numbers. We have already demonstrated the horizontal and vertical cutoff in a single mirror schlieren setup using the LAKE approach \cite{kumar2026using}. However, switching between these two was still mechanical, involving sweeping the LED image from position 1 to position 2 via the tilt and yaw adjustments of the concave mirror (see Fig. \ref{fig:2}a). This switching can be done quickly only if the light source image can move from position 1 to position 2 in a short amount of time. Similarly, both horizontal and vertical cutoffs can be produced simultaneously if the LED image can be present in both positions within a single camera frame. Figure \ref{fig:2}b-\ref{fig:2}c depicts the relationship between lens aperture size $D = 2R$, and the distance between LED placements $d$ as $D = 2R = d\sqrt 2$ which can be rearranged as 
\begin{equation}
fN = f/(d\sqrt2)
\label{eq:1}
\end{equation} where $fN$ represents the lens aperture F-number. A typical camera lens with a focal length of $f = 35$ mm and an F-number value of $fN = 8$ provides $d \simeq 3$ mm. This separation, although small, is easily achievable using surface-mount-type small LEDs.

\begin{figure}[!b]
\centering
\includegraphics[width=0.95\linewidth]{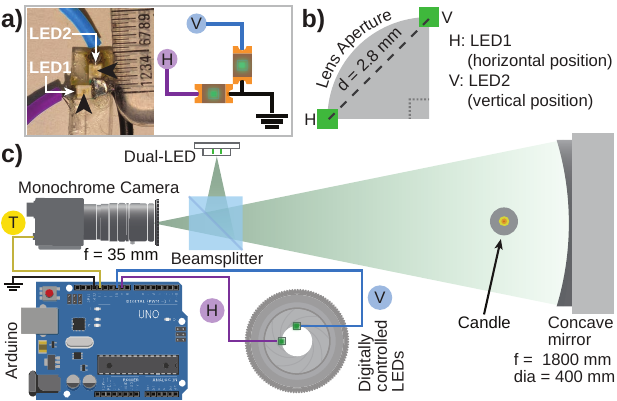}
\caption{Dual LED setup for bidirectional schlieren. a) Two surface mount LEDs are bonded together in an L-shape and wired for separate control. b) Schematic showing the LED separation of 2.8 mm. c) Schematic of the entire schlieren imaging system, which uses the dual-LED configuration in the LAKE method and quick digital control. A beamsplitter aids in maintaining a coincident single-mirror schlieren geometry. The Arduino UNO board provides control signals for the LEDs and the camera trigger.}
\label{fig:3}
\end{figure}

\section{Experiments}
We created a dual-LED illumination arrangement by gluing two surface-mount LEDs (80 mW, Green, 1206 SM-LED) in an L-shaped geometry (see Fig. \ref{fig:3}a). The LEDs were soldered in a common cathode arrangement, resulting in a common ground and two independent anode wires. These LEDs are 3.2 mm $\times$ 1.6 mm small with a tiny 100 $\mu$m emission center (see black arrows in Fig. \ref{fig:3}a). Thus, they do not need pinholes to behave like a point-like light source. The LED emission centers were measured to be 2.8 mm apart (see Fig. \ref{fig:3}b). For a 35 mm focal length lens, this corresponds to an aperture F-number of $fN \approx 8.84$. To match the odd F-number, we used a lens with continuous manual aperture control. Figure \ref{fig:3}c depicts the schematics of the schlieren imaging setup with the dual-LED light source. We employ a large-aperture parabolic concave mirror (Aperture Telescopes, diameter = 400 mm, focal length = 1800 mm). The dual-LED assembly is located in the mirror's center of curvature plane. The light from the dual-LED assembly reflects off a 50:50 beam splitter, illuminating the mirror. The mirror reflected light focuses back on the same dual-LED plane. The beam splitter allows us to access this plane from the transmission side as well. A machine vision camera (Blackfly S USB3: Sony IMX287, 0.4 MP, Mono) equipped with a 35 mm manual lens and mounted on a 3-axis translation stage aids in imaging. The camera assembly is adjusted to align the lens aperture to the dual-LED assumbly. This aligned lens aperture provides a simultaneous horizontal cutoff for the first LED and a vertical cutoff for the second (see Fig. \ref{fig:3}c). To power the LEDs and synchronize the camera frames, we connected the LED pins and camera trigger pin to various output pins on a programmable Arduino UNO board. We now describe various flow visualization experiments using this setup.  

\begin{figure}[ht]
\centering
\includegraphics[width=0.95\linewidth]{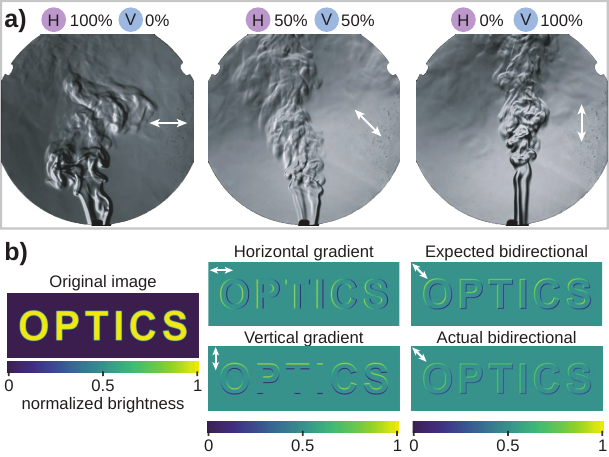}
\caption{Controlling the relative strengths of horizontal and vertical schlieren sensitivity. a) Images with 100\% horizontal, 50\% horizontal + 50\% vertical, and 100\% vertical schlieren sensitivity. White arrows represent perceived schlieren sensitivity orientations. These three images are samples from a sequence of frames taken as the horizontal and vertical LED intensities were gradually reduced and increased, respectively. Visualization 1 depicts the entire sequence. b) Simulating the impact of an equal weighted sum of horizontal and vertical LEDs, resulting in a bidirectional schlieren with reduced sensitivity.}
\label{fig:4}
\end{figure}

The digitally controlled dual LEDs and camera provide complete control over LED brightness and switchability. We use two different approaches for dual LEDs and camera control. In the first approach, we use the Arduino board's Pulse Width Modulation (PWM) signal output to regulate the brightness of LEDs. In the second approach, we use the digital TTL signal output to control LEDs and camera timings (\textbf{Appendix A}). In the first approach, the control sequence gradually reduces the intensity of LED1 from 100\% to 0\%. During the same period, the strength of LED2 increases from 0\% to 100\%, such that the entire sum of both LED strengths equals 100\%. We visualized hot air rising from a burning candle while running this control sequence. This sequence (see \textbf{Visualization 1}) ran for 5 seconds, capturing 264 frames with a 10 ms exposure time and $\sim$52 fps acquisition rate. Figure \ref{fig:4}a depicts three frames from the recorded sequence: the first from the beginning, the second from halfway, and the third from the end. Clearly, the first image is the result of a horizontal cutoff, whereas the third image is due to a vertical cutoff. The middle image, which includes equal contributions from the horizontal and vertical cutoffs, appears to provide a bidirectional schlieren image of the flow. A closer look at the image reveals that the middle image corresponds to a cutoff at a 45 $^\circ$ angle. This ability to change cutoff orientation simply by manipulating the respective strengths of the horizontal and vertical LEDs demonstrates that the addition of cutoffs follows a vector-like sum, with the length of the vector representing the strength of the LED and the direction of the vector representing the cutoff's orientation. However, the similarity to vector sum ends at resultant vector orientation and does not include resultant vector length considerations. The longer the resultant vector, the better the schlieren contrast or sensitivity should be, but the actual outcome is a decrease in contrast. To comprehend this behavior, we conduct a simulation-based investigation, as illustrated in Fig. \ref{fig:4}b. An illustration document with the text "OPTICS" was made using typographic software and saved as an image. The edges of this binary text image were somewhat smoothed using a Gaussian blur filter (kernel size: 5$\times$5). The resulting image serves as our "original image" that represents a density profile. We next extract the mathematical gradient of the original image along the horizontal and vertical orientations, normalize them based on the maximum gradient value, and plot them (see middle column of Fig. \ref{fig:4}b). These gradients optically represent the horizontal and vertical schlieren cutoffs, respectively. Notice how the vertical borders disappear in the horizontal gradient image and vice versa. This behavior is identical to the schlieren image obtained with a knife-edge cutoff. Adding both gradients with 50\% weights should create a bidirectional schlieren image with equal or higher sensitivity (upper right of Fig. \ref{fig:4}b). However, the actual sum produces an image with lesser contrast (bottom right of Fig. \ref{fig:4}b). This decrease in contrast is statistically understandable because of the signal averaging effect. This exact same pattern occurs during schlieren imaging, where the sum of both horizontal and vertical cutoffs results in a decrease in the contrast of the resulting schlieren image. The basic mathematical treatment can also describe the cutoff orientation during the sum (shown by white arrows in Fig. \ref{fig:4}b).

\begin{figure}[!b]
\centering
\includegraphics[width=0.9\linewidth]{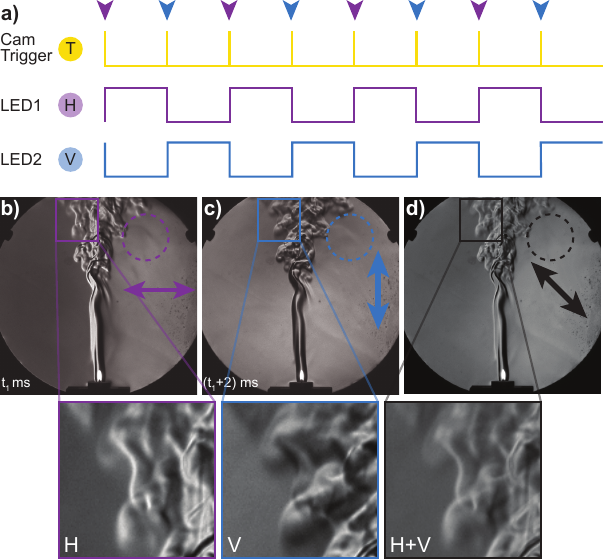}
\caption{Bidirectional schlieren using fast digital switching. a) Synchronized digital control signals for the camera and LEDs. Two adjoining camera triggers are separated by 2 ms, corresponding to 500 fps. b) Schlieren image recorded only with the horizontal LED turned on. c) Schlieren image captured after 2 ms, with only the vertical LED turned on. d) Computational averaging of the previous two frames for bidirectional schlieren. Arrows indicate the schlieren sensitivity orientations. Inset images show that the bidirectional contrast image retains high contrast features along both horizontal and vertical sensitivities. The circled regions indicate the removal of low-contrast flow patterns.}
\label{fig:5}
\end{figure}

The above LED modulation technique produces a sum of horizontal and vertical cutoff images, not a real bidirectional schlieren image. A true bidirectional schlieren with a monochrome camera is only possible when horizontal and vertical cutoff images are isolated from one another. In the past, this was done by spatially moving both horizontal and vertical components on the camera sensor \cite{ chaloupka2014color}. The spatially displaced acquisition makes poor use of the available camera pixels and adds complexity to image registration processes. We consequently use a temporal multiplexing approach in which each camera frame alternates between the horizontal and vertical cutoffs. Figure \ref{fig:5}a depicts the relevant signals for controlling the camera trigger and both LEDs. We recorded 550 frames at a 600 $\mu$s exposure time and 500 fps rate to see hot air rising from a burning candle. The image sequence is presented in \textbf{Visualization 2}. Figures \ref{fig:5}b and \ref{fig:5}c depict two nearby frames from this sequence. These two frames are acquired only 2 ms apart and can be simply combined with a desired weight in post-processing. Figure \ref{fig:5}d illustrates an equal weighted sum of both frames. The inset images depict a zoomed-in view of a flow region, demonstrating how the sum image preserves both horizontal and vertical flow boundaries. The dotted circle shows a region with low-contrast density gradients. These gradients disappear in the sum image, implying a decline in schlieren sensitivity. 

\begin{figure}[!t]
\centering
\includegraphics[width=0.9\linewidth]{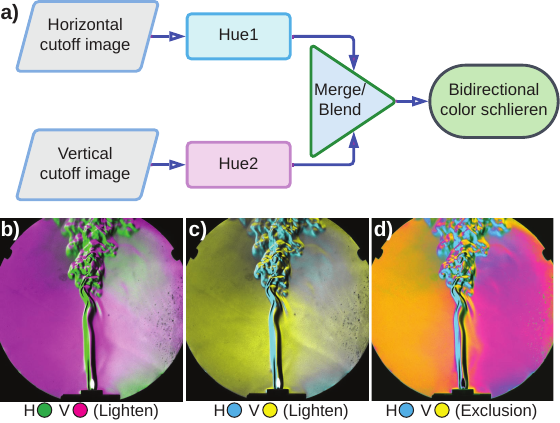}
\caption{Bidirectional color schlieren using monochrome images. a) A flowchart detailing the processes involved in creating a bidirectional color schlieren image from a pair of horizontal and vertical contrast schlieren images. b)-d) Bidirectional color schlieren images created using two hues and one blending options. The choices for complementary and secondary colors with 'lighten' and 'exclusion' blending are presented.}
\label{fig:6}
\end{figure}

An alternating horizontal and vertical cutoff acquisition allows for several postprocessing operations on the image sequence, resulting in a colored bidirectional schlieren image sequence. Figure \ref{fig:6}a depicts the flow chart for this approach. It is feasible to assign any hue and then merge the images using a blending mode of our choice (\textbf{Appendix B}). Figure \ref{fig:6}b-d depicts three examples of hue and blending mode choices. In the first image, we utilize complementary colors for hue and a "lighten" blending setting. In the second image, we use two secondary colors as hues with a "lighten" blending style. In the final image, we use the same two secondary colors for hues while changing the blending mode to "exclusion." Changing hue enables control over color contrast, and switching from the "lighten" blending mode to the "exclusion" mode allows for the appearance of additional hues (magenta and green in our case), which highlights the regions where both horizontal and vertical density gradients are high and low, respectively. \textbf{Visualization 2} displays a side-by-side comparison of the image sequences produced by selecting these hue and blending modes. 

\section{Conclusion}
In conclusion, using dual-LED illumination with a lens aperture as a knife edge, we have demonstrated the control over schlieren cutoff orientation and bidirectional schlieren imaging. We also demonstrated that postprocessing crosstalk-free horizontal and vertical density gradient monochrome image pairs produces a colored bidirectional schlieren image. Despite the employment of two LED sources, we refer to our configuration as a coincident single mirror schlieren arrangement since the distance between the source LEDs (d = 2.8 mm) is substantially smaller than the radius of curvature of the mirror (2f = 3600 mm). The monochrome camera used here is preferred because it lacks RGB filters, which allows it to outperform RGB color camera sensors in terms of quantum efficiency and finer feature resolution. The PWM modulation approach currently utilized to change LED brightness has a significant disadvantage in that it fails when the camera frame rate is comparable to the PWM modulation frequency (490/980 Hz). Thus, at high frame rates, LED modulation must be performed with an actual analog LED driver. Surface-mount LEDs are bright enough to record well-exposed schlieren images at a 600 $\mu$s exposure time, making our approach easily implementable with a few KHz high-frame-rate cameras. We are confident that our approach can be extended to a quantitative measurement of density in flow, resulting in many insightful flow measurements and studies \cite{tp2026study,Elsinga2004Assessment}. 



\begin{backmatter}
\bmsection{Funding} Startup Research Grants from Indian Institute of Technology Delhi and Anusandhan National Research Foundation (ANRF).

\bmsection{Disclosures} A patent application has been filed for the approach presented here. 

\bmsection{Data availability} Data underlying the results presented in this paper are available in ref. \cite{kumar_2026_22301030}.

\bmsection{Supplemental document}
See Supplement 1.
\end{backmatter}

\bibliography{references}
\bibliographyfullrefs{references}

\clearpage
\onecolumn
\section{Appendix}

\subsection{Programming Arduino UNO for LED and camera control} 

We programmed an Arduino UNO board to control LEDs and the camera. We used two different approaches for this control. The first approach uses the PWM output to control LEDs brightness, and the second approach uses the TTL output to synchronize LEDs on/off state and camera trigger. Below we describe both of these approaches.

\subsubsection{PWM output for LEDs brightness control}

Arduino UNO doesn't have a true analog output, but its PWM signal can be used to mimic analog-like behavior to control LED brightness. We start by defining two PWM output pins.

\begin{lstlisting}[numbers=none, caption={}, label={code:1}]
const int PIN_A = 9;  // Define PWM-capable output pins 
const int PIN_B = 10; // Define PWM-capable output pins 
\end{lstlisting}

\noindent We run a serial port to enable user inputs:

\begin{lstlisting}[numbers=none, caption={}, label={code:2}]
Serial.begin(9600); 
Serial.println("Thank you for using this code to control relative strength between two LEDs using PWM.");
Serial.println("Please connect LED1 to Pin 9 and LED2 to Pin 10.");
Serial.println("Enter LED1 brightness percentage (1-100):");
\end{lstlisting}

\noindent We define a simple function to update LED brightness as per the user inputs

\begin{lstlisting}[numbers=none, caption={}, label={code:3}]
// Function to convert 0-100% into 0-255 PWM values and update LEDs
void updateLEDs(int percentage) {
  // map(value, fromLow, fromHigh, toLow, toHigh)
  int pwmA = map(percentage, 0, 100, 0, 255);
  int pwmB = map(100 - percentage, 0, 100, 0, 255);
  analogWrite(PIN_A, pwmA);
  analogWrite(PIN_B, pwmB);
}
\end{lstlisting}

\noindent A user input between 1 and 100 updates both LEDs with a fixed brightness corresponding to a cutoff orientation between 0$^\circ$ and 90$^\circ$. In order to run an automated sequence where LED1 brightness goes from 100\% to 0\% while LED2 goes from 0\% to 100\%, we define another function. This function runs in an angle-dependent sinusoidal modulation scheme as shown below. We have used a $\sin^2$ modulation scheme because it is the squared values of $\sin$ and $\cos$ that add up to unity (or 100\%).

\begin{lstlisting}[numbers=none, caption={}, label={code:4}]
// Function to provide a 5-second sin^2 and cos^2 change to LEDs
void transitionLEDsSinusoidal() {  
  // There are 256 steps (0-255) for covering 90 degree angle (0-90)
  for (unsigned long step = 0; step <= 255; step++) {
    // 1. Calculate the progression angle from 0 to 90 degrees (expressed in radians)
    // Angle goes from 0 to PI/2 radians over the course
    float angleRad = ( (float)step / 255.0 ) * (PI / 2.0);

    // 2. Compute the sinusoidal and convert to pwm value by * 255  
    float sinSquared = sin(angleRad) * sin(angleRad) * 255.0;
    int pwmA = ceil(sinSquared); 
    int pwmB = 255 - pwmA;  

    // 3. Push the % brightness to LEDs
    analogWrite(PIN_A, pwmA);
    analogWrite(PIN_B, pwmB);

    delay(20); // Total transition duration becones 255 * delaytime 
  }
  delay(100);
}
\end{lstlisting}

\noindent The complete code is available in ref. \cite{kumar_2026_22301030}.

\subsection{TTL signals for LED1, LED2 and Camera synchronization}

\noindent In this scheme the Arduino UNO pins are assigned as digital output pins, and their timing is controlled to maintain a 500 fps camera trigger with alternating LED1 and LED2 on times. 
\bigskip

\noindent The pin assignment and initial setup looks like:
\begin{lstlisting}[numbers=none, caption={}, label={code:5}]
const int PIN_A = 9;  // LED1
const int PIN_B = 10; // LED2
const int PIN_T = 13; // Cam Trigger

void setup() {
  pinMode(PIN_A, OUTPUT);
  pinMode(PIN_B, OUTPUT);
  pinMode(PIN_T, OUTPUT);
  
  // Initial state setup
  digitalWrite(PIN_A, HIGH);
  digitalWrite(PIN_B, LOW);
  digitalWrite(PIN_T, LOW);
}

\end{lstlisting}

\noindent The UNO board now has to keep track of the elapsed time and switch between two states:
\begin{lstlisting}[numbers=none, caption={}, label={code:6}]
// --- State 1: Horizontal ---
digitalWrite(PIN_A, HIGH);
digitalWrite(PIN_B, LOW);
digitalWrite(PIN_T, HIGH); //trigger cam post state change

// --- State 2: Vertical ---
digitalWrite(PIN_A, LOW);
digitalWrite(PIN_B, HIGH);
digitalWrite(PIN_T, HIGH); //trigger cam post state change
\end{lstlisting}

\noindent The time is tracked in microseconds via \lstinline|micros()|, and a state change is requested when a duration of 2000 $\mu$s elapses since the last state change. This duration corresponds to 500 frames-per-second. 

\smallskip

\noindent The complete code is available in ref. \cite{kumar_2026_22301030}.

\bigskip
\subsection{Image blending modes}

A standard image editing software provides multiple blending modes between two images. We use Adobe Photoshop \cite{adobe_photoshop} for our images and DaVinci Resolve \cite{davinci_resolve} for our image sequences (videos). The resultant still images are given in the main text (see \textbf{Fig. 6}), and the movie is available as \textbf{Visualization 2}. Both 'lighten' and 'exclusion' blending modes operate on a pixel-by-pixel and channel-by-channel basis (red, green, and blue normalized between 0.0 and 1.0), but they serve entirely different visual functions. Below is a brief description of these two.

\begin{figure}[!t]
\centering
\includegraphics[width=0.99\linewidth]{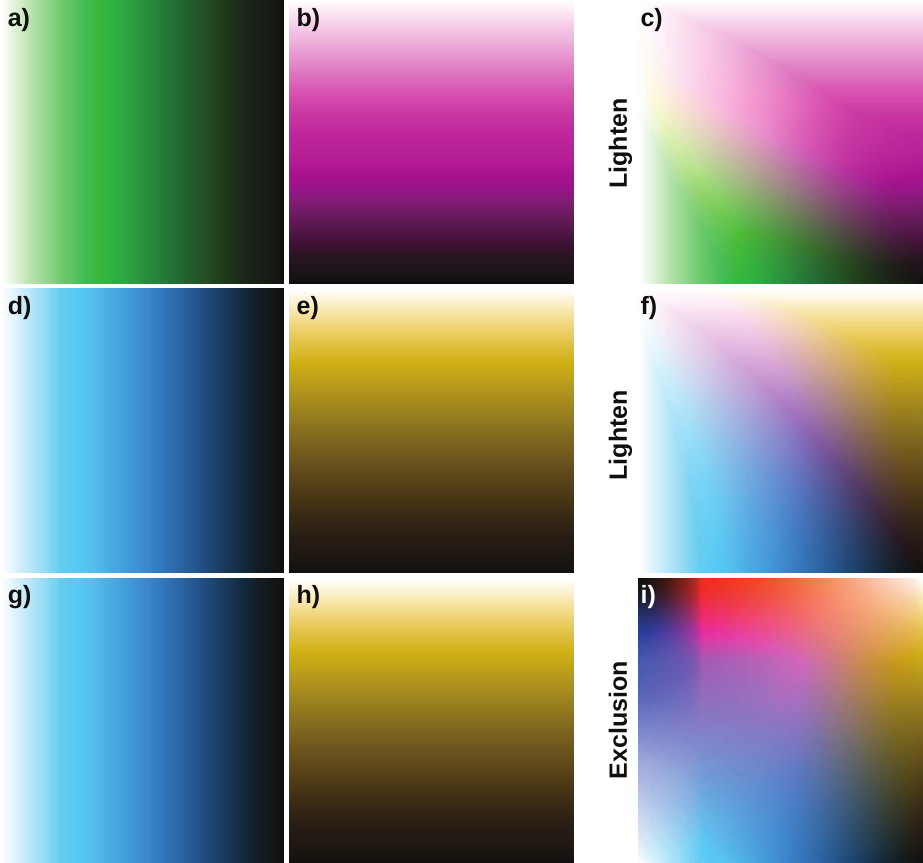}
\caption{Sample for image blending via lighten and exclusion modes. a-b) Input images showing gradual ramp of color Green along horizontal and color Magenta along vertical directions. c) Lighten blending of these two images in a-b. d-e) Input images showing gradual ramp of color Cyan along horizontal and color Yellow along vertical directions. f) Lighten blending of these two images in d-e. g-h) Input images showing gradual ramp of color Cyan along horizontal and color Yellow along vertical directions. i) Exclusion blending of these two images in g-h. }
\label{fig:S1}
\end{figure}

\subsubsection{Lighten}
The lighten mode keeps the lighter color for each RGB channel and discards the darker one. The output image is always as bright as or brighter than both source images. The mathematical relation is given by $\text{Result} = \max(A, B)$ for each channel ($R$, $G$, $B$). The algorithm \ref{alg:lighten} details this blending mode.

\begin{algorithm}
\caption{Lighten Blending}\label{alg:lighten}
\begin{algorithmic}[1]
\Procedure{LightenBlend}{$I_A, I_B$}\Comment{Blends images $I_A$ and $I_B$ using Lighten mode}
\State $W \gets \text{width}(I_A)$
\State $H \gets \text{height}(I_A)$
\For{$x \gets 1 \text{ to } W$}
    \For{$y \gets 1 \text{ to } H$}
        \For{$c \in \{R, G, B\}$}
            \State $A_c \gets I_A[x, y, c]$
            \State $B_c \gets I_B[x, y, c]$
            \State $I_{\text{out}}[x, y, c] \gets \max(A_c, B_c)$\Comment{Keep the lighter channel value}
        \EndFor
    \EndFor
\EndFor
\State \textbf{return} $I_{\text{out}}$
\EndProcedure
\end{algorithmic}
\end{algorithm}

\begin{itemize}
    \item If Pixel $A$ is pure Red $(1, 0, 0)$ and Pixel $B$ is pure Green $(0, 1, 0)$, the result takes the maximum of each channel: $(1, 1, 0)$, which yields Yellow.
    \item Black $(0,0,0)$ is neutral: Blending any color with black leaves the color unchanged because any value is greater than or equal to $0$.
    \item White $(1,1,1)$ dominates: Blending any color with white results in white because no channel value can exceed $1$.
\end{itemize}
\begin{algorithm}
\caption{Exclusion Blending}\label{alg:exclusion}
\begin{algorithmic}[1]
\Procedure{ExclusionBlend}{$I_A, I_B$}\Comment{Blends images $I_A$ and $I_B$ using Exclusion mode}
\State $W \gets \text{width}(I_A)$
\State $H \gets \text{height}(I_A)$
\For{$x \gets 1 \text{ to } W$}
    \For{$y \gets 1 \text{ to } H$}
        \For{$c \in \{R, G, B\}$}\Comment{Normalized color channels $c \in [0, 1]$}
            \State $A_c \gets I_A[x, y, c]$
            \State $B_c \gets I_B[x, y, c]$
            \State $I_{\text{out}}[x, y, c] \gets A_c + B_c - 2 \cdot A_c \cdot B_c$\Comment{Inversion proportional to brightness}
        \EndFor
    \EndFor
\EndFor
\State \textbf{return} $I_{\text{out}}$
\EndProcedure
\end{algorithmic}
\end{algorithm}
\subsubsection{Exclusion}
The exclusion mode inverts light regions while leaves dark regions untouched.  creates a low-contrast, inverted visual effect. Mathematically exclusion mode produces $\text{Result} = A + B - 2AB$, where $A$ and $B$ represent input pixels for both source images given in three channels ($R$, $G$, $B$). The algorithm \ref{alg:exclusion} details this blending mode.

\begin{itemize}
    \item White $(1,1,1)$ inverts: If $B = 1$, then $A + 1 - 2A = 1 - A$. White completely inverts the colors of layer $A$.
    
    \item Black $(0,0,0)$ leaves unchanged: If $B = 0$, $A + 0 - 0 = A$.
    
    \item Mid-grays $(0.5)$ produce 50\% gray: If $B = 0.5$, $A + 0.5 - A = 0.5$, neutralizing color contrast and creating a flat gray tone regardless of $A$'s value.
\end{itemize}

\subsubsection{Blending example}
A visual example of the two blending modes with the hue combinations encountered in the main text is presented here in Fig. \ref{fig:S1}. Clearly, the exclusion blending mode provides a higher number of hues in the output image. However, the hues produced by lighten blending mode are more consistent and visually decipherable.

\end{document}